\documentclass[usenatbib,usegraphicx,letterpaper]{mn2e}
\usepackage[totalwidth=480pt,totalheight=680pt]{geometry}

\usepackage{amssymb}
\usepackage{epsfig}
\usepackage{amsmath}
\usepackage{color}

\newcommand{\unit}[1]{\mathrm{#1}}

\newcommand{\mpc}{\unit{Mpc}}

\newcommand{\hmpc}{\mathrm{Mpc}/h}
\newcommand{\hgpc}{\mathrm{Gpc}/h}
\newcommand{\hkpc}{\mathrm{kpc}/h}
\newcommand{\hmsun}{M_{\odot}/h}

\newcommand{\rvir}{R_\mathrm{vir}}

\newcommand{\msun}{\mathrm{M}_{\odot}}

\newcommand{\lcdm}{\Lambda\mathrm{CDM}}
\newcommand{\mvir}{M_\mathrm{vir}}

\newcommand{\vrad}{V_{\mathrm{rad}}}

\newcommand{\sigrad}{\sigma_{V}^{\mathrm{rad}}}
\newcommand{\siglos}{\sigma_{V}^{\mathrm{los}}}

\newcommand{\ben}{\begin{enumerate}}
\newcommand{\een}{\end{enumerate}}

\usepackage{epsfig}  \usepackage{graphicx}   \usepackage{rotating}

\begin{document}

\title[Assembly Bias \& Cluster RSD]
{Assembly Bias \& Redshift-Space Distortions:  \\ Impact on cluster dynamics tests of general relativity}

\author[Andrew P. Hearin]
{Andrew~P.~Hearin\\
Yale Center for Astronomy \& Astrophysics, Yale University, New Haven, CT
}

\maketitle

\begin{abstract}
The redshift-space distortion (RSD) of galaxies surrounding massive clusters is emerging as a promising testbed for theories of modified gravity. Conventional applications of this method rely upon the assumption that the velocity field in the cluster environment is uniquely determined by the cluster mass profile. Yet, real dark matter halos in N-body simulations are known to violate the assumption that virial mass determines the configuration space distribution, an effect known as {\em assembly bias}. In this Letter, I show that assembly bias in simulated dark matter halos also manifests in velocity space. In the $1-10$ $\mpc$ environment surrounding a cluster, high-concentration ``tracer'' halos exhibit a $10-20\%$ larger pairwise-velocity dispersion profile relative to low-concentration tracer halos of the same mass. This difference is comparable to the size of the RSD signal predicted by $f(R)$ models designed to account for the cosmic acceleration. I use the age matching technique to study how color-selection effects may influence the cluster RSD signal, finding a $\sim10\%$ effect due to redder satellites preferentially occupying higher mass halos, and a $\sim5\%$ effect due to assembly-biased colors of centrals. In order to use cluster RSD measurements to robustly constrain modified gravity, we likely will need to develop empirical galaxy formation models more sophisticated than any in the current literature.
\end{abstract}




\section{INTRODUCTION}
\label{sec:intro}

The observed acceleration of the universe at late cosmological times has been one of the major puzzles of fundamental physics since its discovery. In the standard cosmological model, $\lcdm,$ the cosmic acceleration is accounted for by dark energy, a mysterious, perfectly homogeneous cosmic fluid with a constant equation of state parameter $w=-1.$ 
Alternatively, the cosmic acceleration could instead be driven by modifications to general relativity on cosmological scales. The quest to uncover the physical nature of the acceleration is a primary science target of numerous current and planned galaxy surveys.

Observations of the redshift-space distortions of galaxies (RSD) offer a promising means to constrain models of modified gravity \citep{zhang_etal07,linder08,reyes_etal10,jennings_etal12,reid_etal14,beutler_etal14}. In particular, the velocity structure surrounding massive galaxy clusters is emerging as one of the most sensitive large-scale structure probes of departures from general relativity \citep{schmidt10,lam_etal12,lam_etal13,zu_weinberg13,zu_etal13}. Understanding the infall region of a massive cluster requires making predictions for the velocity field deep into the nonlinear regime, for which the halo model is the premier theoretical tool \citep{seljak01,white01,sheth01,kang_etal02,tinker_etal06,tinker07}. 

The baseline proving grounds for any precision model of velocity structure are cosmological N-body simulations. In particular, for models attempting to predict the velocity field on the highly nonlinear scales of $\sim1-10\mpc,$ it is commonplace to evaluate a model's success by its ability to accurately predict the RSD signal of interest exhibited by host halos selected by mass. For example, a velocity structure model may be deemed accurate if the distribution of pairwise-velocities of dark matter halos is accurately predicted as a function the masses and spatial separation of the halo pairs. 

There are many reasons why the above class of calibration tests can only be considered a prerequisite to the development of a complete model of RSD suitable to potentially falsify general relativity with galaxy redshift surveys. Galaxies are biased tracers of dark matter halos, and the host halo pairwise-velocity field depends on halo mass \citep{tinker07}. Another contribution to the pairwise-velocity field comes from the velocity dispersion of satellite galaxies orbiting within the potential well of their host halo \citep{hikage_yamamoto13}, an effect with a pronounced halo mass-dependence \citep{more09b,wojtak_mamon12}. These effects naturally vary with the selection function of the galaxy sample, and so any model for the velocity field  must be flexible enough to encapsulate variations due to the manner by which galaxies weight halos by mass. 

Although mass is the dominant variable determining the phase space distribution of dark matter halos, it is by now firmly established that knowledge of halo mass alone is insufficient to characterize halo clustering statistics, an effect that goes by the catch-all term {\em halo assembly bias} \citep{gao_etal05,wechsler06,wang_mo_jing07,gao_white07,dalal_etal08,lacerna11}. In addition to configuration space-based statistics such as two-point clustering,  the intra-halo velocity structure of both dark matter particles \citep{faltenbacher_white10} and orbiting subhalos \citep{faltenbacher10} has been shown to depend upon host halo properties besides mass. However, halo assembly bias in the two-halo regime of the velocity field has received considerably less attention. 

If the statistical connection between halos and their resident galaxies depends upon properties besides halo mass, then the galaxy population will inherit the assembly bias of its parent halos. Such a population  is said to exhibit {\em galaxy assembly bias}. It has recently been shown in \citet{zentner_etal13} that galaxy assembly bias can lead to systematic errors that seriously threaten the program to use conventional implementations of the halo model to interpret observations of projected two-point clustering. The findings in \citet{zentner_etal13} are based on an empirical model of galaxy formation (described in \S\ref{sec:sims}), though similar conclusions apply to results based on semi-analytic models \citep[e.g.,][]{croton_etal07,zu_etal08}. 

Motivated by these findings, in this {\em Letter} I take a first look at how color-dependence of the galaxy selection function may influence the  redshift-space distortion signal observed in the neighborhood of massive galaxy clusters. I describe the simulations and mock catalogs I use throughout the paper in \S~\ref{sec:sims}.  I present my results in \S~\ref{sec:results}, and discuss their implications in \S~\ref{sec:conclusions}. 

\section{SIMULATION \& MOCK CATALOGS}
\label{sec:sims}

To study the velocity field surrounding massive halos, I use the collisionless N-body Multidark simulation \citep{riebe_etal11}.  
The Multidark cosmological parameters are based on WMAP5 \citep{komatsu09a}, and 
the simulation was run with $2048^3$
particles of mass $\mathrm{m_p}\approx8.7\times10^{9}\hmsun$ in a $1\hgpc$ periodic box. 
In all the results in this paper, I use ROCKSTAR-based halo catalogs
\citep{behroozi_rockstar11, behroozi_trees13}, publicly available at {\tt
  http://hipacc.ucsc.edu/Bolshoi/MergerTrees.html}. 

In addition to the Multidark halo catalogs, I also study cluster velocity structure using a mock galaxy catalog based on abundance matching \citep{kravtsov04a,conroy06,behroozi10} and 
age matching \citep{HW13a}. In particular, I use the catalog based on $M_r$ luminosity and $g-r$ color, publicly available at 
http://logrus.uchicago.edu/$\sim$aphearin. The value of using the age matching catalogs is that they provide a reasonably realistic representation
 of the galaxy distribution in the local universe \citep{reddick12,hearin_etal12b,hearin_etal13b,watson_etal14}, 
 and they exhibit strong signatures of assembly bias \citep{zentner_etal13}.
In order to parse effects on the velocity field that are due to assembly bias from effects 
 that are purely due to the manner by which halos are weighted by galaxies as a function of halo mass,  
 I additionally use a mock galaxy catalog in which $\mvir$ is the only halo property governing the colors of the halo's galaxies. 
 I construct this ``no-assembly-bias'' mock catalog simply by shuffling the colors between mock galaxies occupying halos of a similar mass, 
 using mass bins $0.1$dex in width, and separately shuffling the colors of centrals and satellites.

\section{RESULTS}
\label{sec:results}

\subsection{Overview of Methodology}
\label{subsec:outline}

In both sections below, I study the pairwise-velocity dispersion of dark matter halos, denoted by $\sigma(R|M_1, M_2),$ where $M_1$ and $M_2$ are the masses of the halo pairs and $R$ is their spatial separation. Motivated by the program to constrain modified gravity described in \citet{schmidt10}, I will focus on the subspace of information provided by the special case where $2$ $\hmpc \lesssim R\lesssim 10$ $\hmpc$ and one of the halo pairs in $\sigma(R|M_1, M_2)$ pertains to a cluster whose mass $M_2\approx10^{14}\hmsun$ is known through independent means, for example through gravitational lensing. 

In all that follows, I keep the cluster sample fixed and study how the cluster velocity dispersion profile depends on the properties of the ``tracer'' $M_1$ halos. In \S\ref{subsec:cdmresults} I show that not only does velocity structure depend on tracer mass $M_1,$ but has an additional dependence on tracer concentration: $$\sigma(R|M_1; M_2=10^{14}\msun)\neq\sigma(R|M_1, c_1; M_2=10^{14}\msun),$$ the manifestation of halo assembly bias in velocity space. Of course, we do not have direct observational access to $M_1,$ and any tracer galaxy sample will occupy a range of halo masses. So in \S\ref{subsec:mockresults} I use the age matching mocks to show how the galaxy occupation statistics of the tracer halos influence mock observations of the velocity dispersion profile, again studying both $M_1-$dependent effects, as well as effects from assembly bias. 


\begin{figure}
\begin{center}
\includegraphics[width=8.5cm]{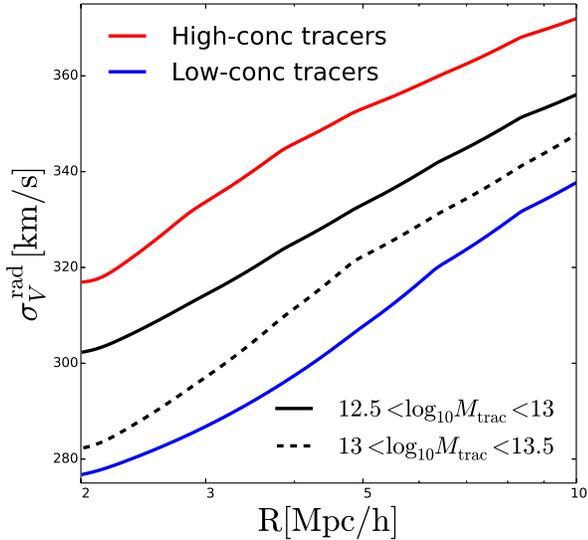}
\caption{ Pairwise radial velocity dispersion profile of tracer halos surrounding cluster-sized halos, plotted as a function 
of cluster-centric distance. Differences between the {\em solid} and {\em dashed black curves} show the dependence of the signal on the mass of the tracer population. {\em Red} and {\em blue curves} show results for a tracer sub-population with high- and low-concentrations for their mass, respectively. The difference between the red and blue curves demonstrates the manifestation of assembly bias in velocity space. Differences in $\sigrad$ are at the $\sim10\%$ level, of order the difference generated by departures from general relativity predicted by $f(R)$ models designed to account for the cosmic acceleration.}
\label{fig:bigbolshoi}
\end{center}
\end{figure}



\begin{figure}
\begin{center}
\includegraphics[width=8.5cm]{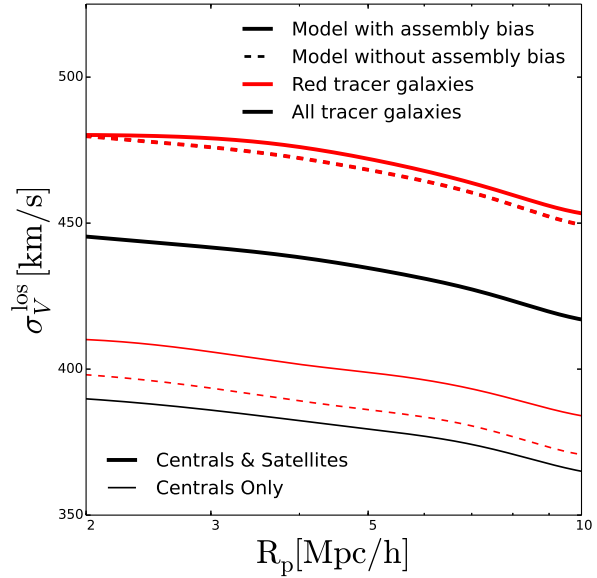}
\caption{Pairwise line-of-sight velocity dispersion profile of mock galaxies surrounding clusters, plotted as a function 
of the projected cluster-centric distance. {\em Lower, thin curves} exclude satellites from the tracer population, {\em upper, thick curves} include the satellite contribution. {\em Black curves} pertain to a tracer population of all galaxies; {\em red curves} show results after applying an additional red sequence cut on the tracers. {\em Solid curves} use the age matching mock, {\em dashed curves} its counterpart in which assembly bias effects have been erased. Differences between the red and black corresponding curves demonstrate the significance of color-selection effects on the cluster RSD signal. Differences between the top and bottom trio of curves indicate the critical role of accurate modeling of satellite velocities. Differences between the red dashed and red solid curves show the influence of assembly bias.}
\label{fig:agm}
\end{center}
\end{figure}


\subsection{Dependence of the Cluster Velocity Field on the Tracer Halo Population}
\label{subsec:cdmresults}

In this section, I demonstrate that the velocity field surrounding massive cluster halos is sensitive to properties of the tracer halo population besides $\mvir.$  I  calculate the velocity structure surrounding Multidark host halos as follows. 
First, I randomly select a sample of $10^{4}$ cluster-sized host halos with mass $10^{13.9}\hmsun<\mvir<10^{14.1}\hmsun,$ 
which has a median $\rvir\approx950\hkpc.$
I select a second sample of host halos of mass $10^{12.5}\hmsun<\mvir<10^{13.5}\hmsun,$ 
which I use as tracers of the velocity field surrounding the cluster-sized halos. In spherical shells surrounding 
each cluster halo, I calculate the radial component of the relative peculiar velocity between the cluster halo and every tracer halo in the shell, denoted by $\vrad.$ 
I then stack all the values of $\vrad$ for every tracer of every cluster, compute the dispersion $\sigrad,$ and plot the result with the black curve in Figure~\ref{fig:bigbolshoi} as a function of the radial distance to the 
mid-point of the spherical shell. 

With the red and blue curves in Figure~\ref{fig:bigbolshoi}, I additionally show 
the velocity structure traced by high- and low-concentration tracer halos in the mass range $10^{12.5}\hmsun<\mvir<10^{13}\hmsun.$ 
To obtain these concentration-selected tracer sub-populations, I bin the tracer halos by mass using $0.02$dex-width bins, and 
split each bin by its median value of NFW concentration, so that the two sub-populations have the same mass function. 

For $R\gtrsim2$ $\hmpc,$ radial velocity dispersion $\sigrad$ decreases with decreasing cluster-centric distance: the relative velocity of halos becomes more coherent as the tracers share an increasing number of large-scale density modes sourcing their velocities \citep{fisher95,reid_white11}. The dependence of $\sigrad$ on tracer-mass reflects the fact that higher mass tracers are less susceptible to tidal forces, so higher mass tracers flow more coherently with the large-scale velocity field \citep[see also the top panels of Figure 6 of][for an alternative demonstration of this point]{tinker07}. The most striking effect shown in Figure~\ref{fig:bigbolshoi} is the trend in tracer concentration. Halos with above-average concentration for their mass exhibit $\sim10-20\%$ higher $\sigrad$ relative to below-average concentration tracers. 

The trend in $\sigrad$ with tracer concentration can be understood in terms of the well-established trend of two-point spatial clustering with concentration. Higher concentration halos cluster more strongly relative to lower concentration halos of the same mass.\footnote{Note that this statement only applies to the mass range relevant to the tracer population shown here; this trend reverses for halos with masses $\mvir\gg M_{*},$ the halo model collapse mass \citep{dalal_etal08}.} Thus at fixed mass, high-concentration halos are found in preferentially over-dense regions. As shown in \citet{tinker07}, the halo-halo pairwise-velocity distribution depends upon large-scale density, even when the masses of the halo pairs are held fixed. Thus sub-selecting tracer halos with high (low) concentration effectively selects pairs in high (low) density environments, naturally giving rise to the character of the trend seen in Figure~\ref{fig:bigbolshoi}. 

\subsection{Dependence of the Cluster Velocity Field on the Tracer Galaxy Population}
\label{subsec:mockresults}

In this section, I use the age matching mock galaxy catalog to translate the results of \S~\ref{subsec:cdmresults} into an observational context. For definiteness, I will focus exclusively on the cluster RSD signal studied in \citet{lam_etal12}, the line-of-sight velocity dispersion profile in the cluster environment. Note, though, that the galaxy occupation principles discussed below apply equally well to alternative statistical quantifications of RSD. 

I begin by selecting all $\sim10^{14}\hmsun$ clusters, defined as galaxy systems residing in host halos with mass $10^{13.75}\hmsun<\mvir<10^{14.25}\hmsun.$ Treating the simulation z-axis as the line-of-sight, I place mock galaxies into redshift-space. For a given tracer population (described below), I select the tracers residing in a cylindrical annulus $4000$ ${\rm km/s}$ in length, and compute the line-of-sight velocity of the tracers in the annulus relative to their associated cluster. Figure~\ref{fig:agm} shows $\siglos,$ the relative line-of-sight velocity dispersion, plotted as a function of the distance to the mid-point of the cylindrical annulus. 

Figure~\ref{fig:agm} shows results for the above calculation pertaining to several different choices for tracer population of mock galaxies. In the {\em bottom trio of thin curves}, I consider an idealized observation in which central galaxy identification is perfect, and all satellites have been excluded from the tracer population. In the {\em top trio of thick curves}, I consider the opposite extreme case where no attempt whatsoever is made to exclude the satellite contribution. In practice, if some BCG-selection and/or isolation criteria is applied to the tracer population, the measured values will lie between the top and bottom trio of curves. {\em Black curves} show results for the full population of $M_r<-19$ tracers, while {\em red curves} pertain to tracers passing an additional red sequence cut, defined by $(g-r) = 0.21 - 0.03M_r.$ {\em Solid red curves} use colors taken directly from the age matching mock, while {\em dashed red curves} pertain to the second mock galaxy catalog in which assembly bias has been erased by the shuffling procedure described in \S~\ref{sec:sims}. 

The first general trend to note is the change in the slope of all curves in Figure~\ref{fig:agm} relative to those in Figure~\ref{fig:bigbolshoi}. This difference can be understood through the characteristic shape of the pairwise-velocity dispersion profile as a function of scale \citep[][Figure 4]{tinker_etal06}. On sufficiently small scales, the dispersion profile turns over and increases with decreasing pair separation due to the mutual gravitational influence of the halo pairs. The radial dispersion does not exhibit this turnover until $R\lesssim2\hmpc,$ but  line-of-sight smearing together with the $\Delta z$ cut result in the turnover occuring at $R_{\rm p}\approx10\hmpc.$

Let us now unpack how the galaxy sample selection function can influence the cluster RSD signal by comparing various pairs of curves in Figure~\ref{fig:agm}. The simplest comparison to understand is the difference between the top and bottom trio of  curves. In all cases, the intra-halo velocity dispersion of satellites significantly enhances $\siglos$ on all scales, as should be expected. 

Next consider the signal exhibited by the idealized centrals-only samples shown with the lower trio of thin curves. At the fainter end of the luminosity function, age matching predicts that red centrals reside in slightly lower mass halos relative to blue centrals of the same brightness. Recall from Figure~\ref{fig:bigbolshoi} that $\sigma_{V}$ decreases with increasing halo mass. In the model with no assembly bias, this is the only operative trend for centrals, and so {\em red} centrals living in less massive halos exhibit a larger $\siglos$ relative to {\em all} centrals: the thin, red dashed curve lies above the thin, solid black curve. In the age matching model that includes assembly bias, red centrals not only live in lower mass halos, but red centrals additionally occupy halos of above-average concentration. This assembly bias effect further boosts the velocity dispersion profile of red sequence centrals: the thin, solid red curve shows a $\sim5\%$ boost relative to the thin, dashed red curve. This assembly bias boost is simply the concentration trend shown with the red and blue curves Figure~\ref{fig:bigbolshoi}, as  manifested in the age matching mock. 

Finally, consider the top trio of thick curves illustrating results for tracer galaxies that include a satellite contribution, beginning by comparing the thick, dashed red and thick, solid black curves. This comparison demonstrates the effect that a red sequence cut has on $\siglos$ for a galaxy population in which color is governed exclusively by virial mass. Any red sequence cut has two, related effects on the satellite population:
\ben
\item At fixed luminosity, redder populations of galaxies have a larger satellite fraction; 
\item redder populations of satellites are found in preferentially more massive host halos. 
\een
Both of these ``$\mvir-$only'' selection effects work together so that a red sequence cut generically boosts $\siglos$ due to an increased contribution from the internal motion of satellites within their host halos. 

Finally, let us consider the role played by assembly bias in the tracer samples that include satellites by examining the thick, solid red curve. In the galaxy sample illustrated by this curve, the color-selection effects on the satellite population discussed in the previous paragraph still apply, but there is an additional effect due to assembly bias in both the centrals and satellites. From the bottom trio of curves, we know that assembly bias in the central population will generically lead to a boost $\siglos$ after applying the red sequence cut. However, as shown in \citet{zentner_etal13}, age matching predicts much weaker assembly bias effects for satellites relative to centrals (see also McEwan et al. 2015, in prep). Thus although assembly bias is still operative in the sample that includes satellites, the ``$\mvir-$only'' effects dominate. For measurements made on tracer samples with no satellite exclusion criteria, age matching  predicts the role played by assembly bias in color-selection effects on $\siglos$ to be limited to $1-2\%,$ as shown by the difference between the thick, solid red and thick, solid dashed curves. 

\section{DISCUSSION}
\label{sec:conclusions}

There is rich information about our cosmology encoded in $P(\Delta V_{12}|R,M_1, M_2),$ the pairwise-velocity distribution of halos \citep{tinker07,reid_etal14}. Recent work has focused on the special case where one of the halo pairs is a cluster with mass that is assumed to be known by lensing \citep[e.g.,][]{lam_etal13,zu_etal13}. Preliminary indications are quite promising, as these authors have shown that  cluster velocity structure can be modeled with reasonably good accuracy, and with a signal that varies at the $10-20\%$-level due to modified gravity effects. 

In practice, we must rely upon galaxy redshift surveys to provide our information about the velocity field, and so regardless of the approach to the RSD modeling, the method must be able to account for the biased manner by which the tracer galaxies occupy halos. The results in \S\ref{subsec:mockresults} based on the age matching mocks show that the effects of the galaxy selection function on the cluster velocity field are roughy as large as the modified gravity signal itself. These findings imply that the program to place precision constraints on general relativity with cluster RSD will need to proceed hand-in-hand with realistic models connecting galaxies to dark matter halos. 

The success of age matching implies it is at least plausible that galaxy color correlates with halo assembly at fixed halo mass. However, traditional implementations of empirical galaxy-halo models such as the Halo Occupation Distribution \citep{berlind02} and Conditional Luminosity Function \citep{yang03} neglect to account for assembly bias; such implementations are unable even in principle to encapsulate the complexity shown in Fig.~\ref{fig:agm}. The semi-analytic approach to the galaxy-halo connection \citep[see, e.g.,][and references therein]{benson12,henrique_etal14} is not limited by such simplifying assumptions, but at present this class of models is too computationally demanding to use in cosmological parameter likelihood analyses. In order to robustly exploit the cluster RSD signal that will be measured with Stage IV dark energy experiments, we likely need to develop a new generation of galaxy-halo models more sophisticated than any in the current literature.

\section*{ACKNOWLEDGEMENTS}
\label{sec:ack}

Thanks to Nikhil Padmanabhan, Andrew Zentner, Beth Reid, Jeremy Tinker, 
and Frank van den Bosch for informative conversations, 
and to Alexie Leauthaud and Fabian Schmidt for comments on an early draft. 
Special thanks to Andrew Wetzel for providing sanity check calculations. 
I thank Blockhead for {\em Carnivores Unite}. 

\bibliography{./grclust.bib}


\end{document}